\documentclass{article}
\usepackage{spconf,amsmath,amssymb,graphicx,booktabs,multirow,hyperref,siunitx}
\usepackage[table]{xcolor}
\graphicspath{{figures/assets/}}
\usepackage{array,tabularx}
\hypersetup{hidelinks}

\definecolor{oursbg}{HTML}{E5E5FF}

\newcommand{\Sone}{\textsc{Stage 1}}
\newcommand{\Stwo}{\textsc{Stage 2}}

\newcommand{\LCE}{\mathcal L_{\mathrm{CE}}}
\newcommand{\Lpres}{\mathcal L_{\mathrm{pres.}}}
\newcommand{\Dsnr}{\Delta_{S}}

\title{Transferable Adversarial Robustness for Speech Foundation Models via Hierarchical Stabilization}

\name{\begin{tabular}{c}
Aref Mousavi$^{1}$, Shahab Sherafat$^{2,*}$, Kiarash Kiani Feriz$^{1,*}$, Amirparsa Safari$^{1,*}$ \\
Raoof Zare Moayedi$^{1}$, Mohammad Hossein Rohban$^{1}$, Mohammad Sabokrou$^{3,4}$
\end{tabular}\thanks{$^{*}$Equal contribution.}}

\address{{\small
$^{1}$Sharif University of Technology \hspace{0.45em}
$^{2}$University of Tehran \hspace{0.45em}
$^{3}$New Uzbekistan University \hspace{0.45em}
$^{4}$Okinawa Institute of Science and Technology}}

\makeatletter
\def\ps@arxivfirst{%
  \def\@oddhead{}\def\@evenhead{}%
  \def\@oddfoot{\hfil\parbox{0.96\textwidth}{\centering\scriptsize
    This work has been submitted to the IEEE for possible publication. Copyright may be transferred without notice, after which this version may no longer be accessible.}\hfil}%
  \let\@evenfoot\@oddfoot}
\makeatother

\begin{document}
\ninept
\maketitle
\thispagestyle{arxivfirst}

\setlength{\abovedisplayskip}{4pt plus 2pt minus 2pt}
\setlength{\belowdisplayskip}{4pt plus 2pt minus 2pt}
\setlength{\abovedisplayshortskip}{2pt plus 1pt}
\setlength{\belowdisplayshortskip}{2pt plus 1pt}

\begin{abstract}
Frozen speech foundation models (SFMs) make downstream adaptation efficient: the backbone can stay fixed while a task learns layer fusion and a lightweight classifier. Full adversarial fine-tuning is a standard route to robustness, but generating adversarial examples and updating the backbone for every task sacrifices that efficiency. We ask whether robustness can instead be learned before future tasks are known. For a frozen backbone and linear classifier, robustness can be understood through the interaction between representation stability and decision-boundary margin. This leads directly to our design: we stabilize representations across the hidden layers, rather than only the final layer, while preserving clean representations; after clean adaptation selects the layer mixture, we keep it fixed and enlarge only the classifier margin, without downstream adversarial examples. We evaluate Wav2Vec2, HuBERT, and WavLM Large on four tasks under adaptive 30 dB attacks. Across 12 backbone--task pairs, hierarchical robustification improves robust accuracy by 46.4 pp, while margin refinement adds 4.0 pp for 1.1 pp of clean accuracy. 
Code and configurations are available at \href{https://github.com/arefmousavi/hierarchical-robust-sfm}{\nolinkurl{github.com/arefmousavi/hierarchical-robust-sfm}}.
\end{abstract}


\begin{figure*}[t]
\centering
\includegraphics[width=\textwidth]{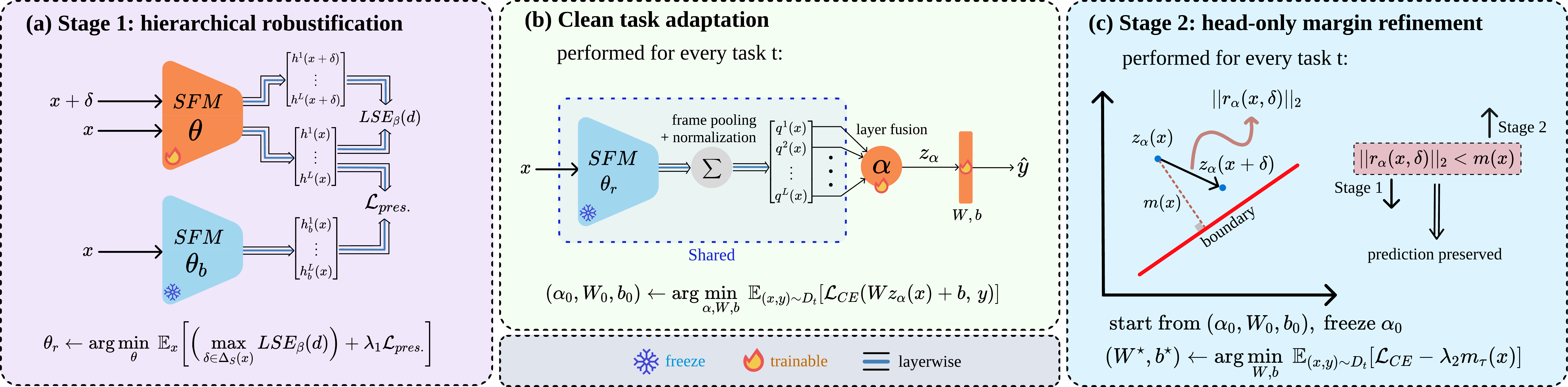}\vspace{-1mm}
\caption{Overview of the proposed framework. \textbf{\Sone{}} robustifies representations across hidden layers before downstream tasks are known, while preserving clean representations. Clean adaptation then learns task-specific convex layer fusion and a linear classifier with the robust backbone frozen. \textbf{\Stwo{}} fixes the backbone and fusion and refines only the classifier using normalized geometric margins.}
\label{fig:method}
\end{figure*}

\section{Introduction}
\label{sec:intro}

Frozen speech foundation models (SFMs) support efficient downstream adaptation because the backbone can remain fixed. In the SUPERB-style classification setting considered here, a task learns a weighted fusion of frozen layer representations together with a lightweight classifier~\cite{yang2021superb}. Large-scale evaluation shows that learnable layer fusion outperforms final-layer-only evaluation in most cases, while preferred depths vary across tasks~\cite{pasad2023layerwise,yang2024largescale}. Full task-specific adversarial fine-tuning is a direct route to robustness, but iterative adversarial-example generation and end-to-end backbone updates are costly and must be repeated for every new task~\cite{madry2018robust,shafahi2020robusttransfer}. We therefore ask whether the efficiency of frozen-backbone adaptation can be retained while robustness is learned before future downstream tasks are known.

This goal requires more than a robust terminal embedding. FARE in Robust CLIP stabilizes the CLIP vision-encoder output so existing downstream systems can reuse the robust encoder without retraining~\cite{radford2021clip,schlarmann2024robustclip}. In our setting, however, the future speech task chooses both a layer mixture and a classifier after backbone robustification. For a frozen backbone and linear classifier, robustness is governed by the interaction between representation stability and decision-boundary margin. Let $z_\alpha(x)$ be the fused representation, $r_\alpha(x,\delta)=z_\alpha(x+\delta)-z_\alpha(x)$ the attack-induced movement, and $m(x)$ the Euclidean distance to the nearest classifier decision boundary. A sufficient condition for preserving the prediction is
\begin{equation}
 \|r_\alpha(x,\delta)\|_2<m(x).
 \label{eq:central}
\end{equation}
This view leads directly to our design. Before tasks are known, we stabilize representations across the hidden layers, rather than only the final layer, while preserving clean representations. After clean adaptation selects the task-specific fusion and classifier, we keep the backbone and fusion fixed and refine only the classifier margin. Keeping the fusion fixed lets the margin objective change decision geometry without changing the representation selected by clean adaptation, and requires no downstream adversarial examples. Fig.~\ref{fig:method} summarizes the resulting pipeline and its movement--margin interpretation.

Prior work establishes that robust pretrained representations can transfer robustness to downstream classifiers and studies how that robustness behaves during adaptation~\cite{shafahi2020robusttransfer,chen2020advssl,nern2023transfer,hua2024roli,mao2023tecoa}. In speech, Wu et al. study attacks when the downstream task, layer-fusion weights, and classifier are unknown to the attacker~\cite{wu2022sslattack}, Olivier et al. demonstrate adversarial transfer across self-supervised speech models~\cite{olivier2022watch}, and Wang et al. study task-specific adversarial fine-tuning for robust ASR~\cite{wang2026vac}. These works do not address the combination studied here: future task-specific layer fusion and classifier are unknown during robustification, representations must therefore be stabilized across layers, and the downstream decision geometry is refined after adaptation, using no adversarial examples.

Our main contributions are: \textbf{(1) Hierarchy-aware adversarial robustification.} We introduce an adversarial objective matched to frozen SFM layer fusion that stabilizes representations across hidden layers while preserving clean representations. \textbf{(2) Movement-margin formulation.} We characterize robustness through the interaction between adversarial representation movement and distance to the nearest classifier boundary, and show that across-layer stabilization bounds this movement for every layer fusion a task may select. \textbf{(3) Clean margin-aware classifier refinement.} We refine only the downstream classifier, with the robust backbone and learned fusion fixed, using no downstream adversarial examples. We evaluate Wav2Vec2, HuBERT, and WavLM Large on four complementary speech classification tasks using adaptive white-box attacks.

\section{Method}
\label{sec:method}

\subsection{Notation and layer-normalized representations}
Let $x$ be a waveform and let $h_\theta^\ell(x)\in\mathbb R^{T_x\times D}$ be the sequence of $T_x$ valid (unpadded) frames that a backbone with parameters $\theta$ produces at layer $\ell$, where $D$ is the hidden dimension and $\ell=1,\dots,L$ indexes the outputs of the $L$ transformer blocks. All $L$ representations of an utterance share the frame count $T_x$. The pretrained backbone is $\theta_b$ and its robustified counterpart $\theta_r$; $\Pi(h)=\frac{1}{T_x}\sum_{t=1}^{T_x}h_t\in\mathbb R^{D}$ denotes mean pooling over valid frames.

Layers cannot be compared on raw distances, because activation scale varies across depth. We therefore fix one scale per layer, $\sigma_\ell^2=\mathbb E_x\big[\frac{1}{T_xD}\|h_{\theta_b}^\ell(x)\|_F^2\big]$, estimated once on held-out clean speech independent of any downstream task, and measure the discrepancy between two frame sequences $a,b\in\mathbb R^{T_x\times D}$ by
\begin{equation}
 \rho_\ell^{2}(a,b)=
 \frac{1}{T_x D\,\sigma_\ell^2}\sum_{t=1}^{T_x}\|a_t-b_t\|_2^2 .
 \label{eq:rho}
\end{equation}
Because the $\sigma_\ell$ are estimated from $\theta_b$ and then held fixed, no method can shrink Eq.~\eqref{eq:rho} by inflating its own normalizer, so layer drift stays comparable across methods. A downstream task consumes the same layers after pooling, centering, and normalization:
\begin{equation}
 q_\ell(x)=\frac{\Pi(h_{\theta_r}^\ell(x))-\mu_\ell}{\sigma_\ell\sqrt D}\in\mathbb R^{D},\qquad
 z_\alpha(x)=\sum_{\ell=1}^{L}\alpha_\ell q_\ell(x),
 \label{eq:q}
\end{equation}
where $\mu_\ell$ is the mean of $\Pi(h_{\theta_r}^\ell(x))$ over a held-out split of the task data $\mathcal D_t$, and the fusion weights satisfy $\alpha_\ell\ge0$ and $\sum_{\ell=1}^{L}\alpha_\ell=1$. A linear classifier $g(z)=Wz+b$, with $W\in\mathbb R^{K\times D}$ and rows $w_1^\top,\dots,w_K^\top$, acts on $z_\alpha$. Because Eq.~\eqref{eq:rho} and Eq.~\eqref{eq:q} share $\sigma_\ell$, robustification and downstream adaptation are expressed in one coordinate system. 
% The centering $\mu_\ell$ is the same for every input, so it cancels when two representations are subtracted: differences of $q_\ell$ are differences of $\Pi(h_{\theta_r}^\ell(x))$ divided by the constant $\sigma_\ell\sqrt D$.

\subsection{\Sone: adversarial stability across layers}
At robustification time the future task, its fusion weights, and its classifier are all unknown, so no layer can be favored in advance. We score the instability of layer $\ell$ by $d_\ell(x,\delta;\theta)=\rho_\ell^{2}(h_\theta^\ell(x),h_\theta^\ell(x+\delta))$ and aggregate $\mathbf d=(d_1,\dots,d_L)$ across depth with the smooth maximum
\begin{equation}
 \operatorname{LSE}_\beta(\mathbf d)=
 \frac{1}{\beta}\log\!\left[\frac{1}{L}\sum_{\ell=1}^{L}e^{\beta d_\ell}\right],
 \label{eq:lse}
\end{equation}
which brackets the worst layer: $\max_\ell d_\ell-\tfrac{1}{\beta}\log L\le\operatorname{LSE}_\beta(\mathbf d)\le\max_\ell d_\ell$, with $\beta\to0$ recovering the layer mean. 
Robustness must not destroy the pretrained features, and $\mathbf d$ can also be driven down by shrinking activations, so we anchor the clean representation to $\theta_b$,
\begin{equation}
 \Lpres(x)=\frac{1}{L}\sum_{\ell=1}^{L}\rho_\ell^{2}\!\left(h_\theta^\ell(x),h_{\theta_b}^\ell(x)\right),
 \label{eq:pres}
\end{equation}
and, using unlabeled task-independent speech, solve
\begin{equation}
 \theta_r=\arg\min_\theta\mathbb E_x\!\left[
 \Big(\max_{\delta\in\Dsnr(x)}\operatorname{LSE}_\beta(\mathbf d)\Big)+\lambda_1\Lpres(x)\right],
 \label{eq:stage1}
\end{equation}
where $\Dsnr(x)$ is the perturbation set of Eq.~\eqref{eq:threat} at $S=30$ dB and the inner maximization is approximated by projected gradient ascent. $\Lpres$ sits outside it because it does not depend on $\delta$.

\textbf{Why the objective must span layers.} Let $\Delta q_\ell=q_\ell(x+\delta)-q_\ell(x)$, so that $r_\alpha(x,\delta)=\sum_{\ell}\alpha_\ell\Delta q_\ell$. Pooling averages over frames, so Jensen's inequality gives $\|\Delta q_\ell\|_2^2\le d_\ell$, both sides at $\theta_r$: a layer's pooled displacement cannot exceed the frame-level quantity \Sone{} minimizes. With the convexity of the mixture and the two-sided bound on $\operatorname{LSE}_\beta$,
\begin{equation}
\begin{aligned}
 \|r_\alpha(x,\delta)\|_2&\le\max_{\ell}\|\Delta q_\ell\|_2\le\big(\max_{\ell}d_\ell\big)^{1/2}\\
 &\le\big(\operatorname{LSE}_\beta(\mathbf d)+\tfrac{1}{\beta}\log L\big)^{1/2}
\end{aligned}
 \label{eq:convexbound}
\end{equation}
for every admissible $\alpha$: \Sone{} controls the movement term of Eq.~\eqref{eq:central} uniformly over the fusions a future task might choose. Minimizing $d_L$ alone leaves the other $d_\ell$ unconstrained, so it bounds $\|r_\alpha\|_2$ only for the mixture that uses layer $L$ by itself.
\begin{table*}[!t]
\newcommand{\tabvalue}[2]{\mbox{{\fontsize{5.5}{6.5}\selectfont #1}\raisebox{-0.6pt}{\textcolor{black!45}{\scalebox{0.70}{\fontsize{6}{7}\selectfont\ensuremath{\pm}#2}}}}}
\newcommand{\tabpair}[4]{\tabvalue{#1}{#2}\hspace{2pt}/\hspace{2pt}\tabvalue{#3}{#4}}
\centering
\caption{Robustness transfer across three backbones and four tasks. Task entries are clean/robust accuracy (CA/RA, \%), reported as mean$\pm$SD over three downstream adaptation seeds. The four overbarred columns report changes relative to Standard+CE, averaged across tasks within each backbone; $\Delta$CA and $\Delta$RA are in percentage points, while $\Delta R_\alpha$ and $\Delta M_{10}$ are defined in Eq.~\eqref{eq:diagnostic_changes}. $\dagger$ indicates use of downstream adversarial examples; Full AT is a task-specific oracle outside the frozen-backbone setting.}
\label{tab:main}
\vspace{1mm}
\fontsize{6}{7}\selectfont
\setlength{\tabcolsep}{2.5pt}
\renewcommand{\arraystretch}{1.0}
\begin{tabularx}{\textwidth}{@{}>{\raggedright\arraybackslash}X *{4}{>{\centering\arraybackslash}p{2.48cm}} *{4}{>{\fontsize{5.5}{6.5}\selectfont}c} @{}}
\toprule
& \multicolumn{4}{c}{\textbf{Clean / robust accuracy (\%)} $\uparrow$}
& $\overline{\Delta R_\alpha}$
& $\overline{\Delta M_{10}}$
& $\overline{\Delta\mathrm{CA}}$
& $\overline{\Delta\mathrm{RA}}$ \\
\cmidrule(lr){2-5}
\noalign{\vskip -1pt}
\multirow{-2}{*}{\textbf{Method}} & \textbf{KS} & \textbf{IC} & \textbf{SID} & \textbf{ER} 
& (\%) $\downarrow$
& ($\times10^{-3}$) $\uparrow$
& (pp) $\uparrow$
& (pp) $\uparrow$ \\
\midrule
% % ------------------------------------------------------------------
% % Wav2Vec2 Large
% % ------------------------------------------------------------------
\multicolumn{9}{@{}l}{\textbf{Wav2Vec2 Large (primary backbone)}} \\
\quad Standard + CE & \tabpair{95.2}{0.56}{ 5.5}{0.96} & \tabpair{94.1}{0.47}{ 2.7}{0.52} & \tabpair{87.6}{0.91}{ 0.5}{0.09} & \tabpair{67.5}{0.52}{ 0.8}{0.18} & 0.0 & 0.0 & 0.0 & 0.0 \\

\quad FARE-style + CE & \tabpair{91.7}{0.19}{39.7}{0.71} & \tabpair{90.5}{0.56}{36.8}{0.63} & \tabpair{83.9}{0.48}{16.0}{0.99} & \tabpair{64.9}{1.48}{ 9.7}{0.95} & -38.3 & 0.5 & -3.4 & 23.2 \\

\quad Terminal-layer + CE & \tabpair{92.5}{0.54}{38.2}{0.75} & \tabpair{91.2}{0.17}{36.1}{1.36} & \tabpair{84.1}{0.92}{15.6}{0.58} & \tabpair{64.8}{0.62}{ 9.8}{1.43} & -37.5 & 2.1 & -3.0 & 22.6 \\

\quad \textbf{\Sone{} + CE} & \tabpair{93.7}{0.56}{62.2}{0.47} & \tabpair{92.8}{0.58}{61.0}{0.45} & \tabpair{84.4}{0.47}{34.8}{0.64} & \tabpair{65.2}{0.73}{26.5}{1.27} & -81.8 & 8.6 & -2.1 & 43.8 \\

\rowcolor{oursbg}
\quad \quad + \textbf{\Stwo{} (ours)} & \tabpair{92.3}{0.35}{67.4}{1.22} & \tabpair{91.6}{0.55}{66.7}{1.15} & \tabpair{83.6}{0.49}{37.7}{0.65} & \tabpair{64.0}{0.38}{29.6}{0.70} & -82.2 & 13.1 & -3.2 & 48.0 \\

\quad \quad + adversarial head training$^\dagger$ & \tabpair{90.4}{0.57}{71.8}{0.53} & \tabpair{89.4}{0.24}{69.0}{0.54} & \tabpair{81.5}{0.68}{39.8}{0.97} & \tabpair{62.1}{0.31}{31.8}{0.60} & -82.6 & 14.0 & -5.3 & 50.7 \\
\midrule
% % ------------------------------------------------------------------
% % HuBERT Large
% % ------------------------------------------------------------------
\multicolumn{9}{@{}l}{\textbf{HuBERT Large}} \\
\quad Standard + CE & \tabpair{94.2}{0.54}{ 5.9}{1.04} & \tabpair{97.0}{0.56}{ 4.6}{0.57} & \tabpair{90.6}{0.58}{ 0.7}{0.13} & \tabpair{68.9}{1.12}{ 0.9}{0.17} & 0.0 & 0.0 & 0.0 & 0.0 \\

\quad FARE-style + CE & \tabpair{91.5}{0.63}{40.6}{0.89} & \tabpair{93.7}{0.61}{43.7}{0.30} & \tabpair{87.3}{1.35}{17.9}{0.70} & \tabpair{66.2}{0.84}{11.3}{0.60} & -40.5 & 3.5 & -3.0 & 25.4 \\

\quad Terminal-layer + CE & \tabpair{90.7}{0.32}{42.1}{0.73} & \tabpair{94.4}{0.53}{42.3}{0.73} & \tabpair{87.0}{0.50}{18.6}{0.35} & \tabpair{66.3}{0.39}{11.4}{1.63} & -41.3 & 3.4 & -3.1 & 25.6 \\

\quad \textbf{\Sone{} + CE} & \tabpair{92.6}{0.49}{64.3}{1.24} & \tabpair{95.9}{0.58}{67.9}{0.54} & \tabpair{87.5}{1.22}{37.1}{1.08} & \tabpair{66.8}{0.81}{29.1}{0.63} & -85.9 & 9.2 & -2.0 & 46.6 \\

\rowcolor{oursbg}
\quad \quad + \textbf{\Stwo{} (ours)} & \tabpair{91.2}{0.54}{69.4}{1.29} & \tabpair{94.5}{0.53}{72.5}{1.36} & \tabpair{86.8}{0.53}{41.5}{0.78} & \tabpair{65.5}{0.33}{32.7}{0.91} & -86.1 & 13.8 & -3.2 & 51.0 \\

\quad \quad + adversarial head training$^\dagger$ & \tabpair{89.4}{0.42}{72.8}{1.32} & \tabpair{92.4}{0.27}{74.2}{0.53} & \tabpair{84.7}{0.81}{43.3}{1.63} & \tabpair{63.6}{0.61}{34.9}{1.18} & -86.0 & 15.1 & -5.2 & 53.3 \\
\midrule
% % ------------------------------------------------------------------
% % WavLM Large
% % ------------------------------------------------------------------
\multicolumn{9}{@{}l}{\textbf{WavLM Large}} \\
\quad Standard + CE & \tabpair{96.1}{0.60}{ 7.5}{1.41} & \tabpair{96.7}{0.58}{ 3.1}{0.09} & \tabpair{94.3}{0.54}{ 0.8}{0.15} & \tabpair{71.2}{1.23}{ 1.3}{0.47} & 0.0 & 0.0 & 0.0 & 0.0 \\

\quad FARE-style + CE & \tabpair{92.7}{0.54}{44.0}{0.33} & \tabpair{93.6}{0.55}{40.4}{0.60} & \tabpair{90.8}{0.39}{20.8}{1.48} & \tabpair{68.6}{0.40}{12.8}{0.58} & -35.7 & 1.5 & -3.2 & 26.3 \\

\quad Terminal-layer + CE & \tabpair{93.5}{0.33}{42.8}{0.22} & \tabpair{93.2}{0.47}{41.9}{0.90} & \tabpair{91.4}{0.75}{19.3}{1.26} & \tabpair{68.5}{1.02}{12.9}{0.49} & -35.6 & 1.7 & -2.9 & 26.1 \\

\quad \textbf{\Sone{} + CE} & \tabpair{94.4}{0.54}{69.1}{1.20} & \tabpair{95.3}{0.37}{65.6}{0.44} & \tabpair{91.3}{0.53}{41.7}{0.76} & \tabpair{68.9}{1.37}{31.5}{1.12} & -80.5 & 5.6 & -2.1 & 48.8 \\

\rowcolor{oursbg}
\quad \quad + \textbf{\Stwo{} (ours)} & \tabpair{93.1}{0.20}{71.6}{1.28} & \tabpair{94.4}{0.58}{71.1}{1.25} & \tabpair{90.9}{1.27}{44.4}{0.76} & \tabpair{67.7}{0.22}{34.0}{0.64} & -80.8 & 9.7 & -3.1 & 52.1 \\

\quad \quad + adversarial head training$^\dagger$ & \tabpair{91.2}{0.29}{75.2}{0.59} & \tabpair{92.0}{0.34}{73.5}{0.66} & \tabpair{88.3}{1.06}{48.0}{0.81} & \tabpair{65.8}{0.77}{36.9}{1.54} & -81.2 & 9.9 & -5.3 & 55.2 \\
\midrule
% % ------------------------------------------------------------------
% % Full downstream AT
% % ------------------------------------------------------------------
\multicolumn{9}{@{}l}{\emph{Full downstream AT oracle -- Wav2Vec2 Large}} \\
\quad Standard + full AT$^\dagger$ & \tabpair{91.0}{1.30}{82.9}{3.09} & \tabpair{90.8}{1.31}{77.5}{3.05} & \tabpair{82.1}{1.11}{53.0}{1.97} & \tabpair{62.6}{1.66}{38.7}{2.49} & -- & -- & -- & -- \\
\bottomrule
\end{tabularx}
\end{table*}

\subsection{Clean adaptation and the movement-margin condition}
For task $t$ we freeze $\theta_r$ and learn the fusion weights and linear classifier from clean labels alone,
\begin{equation}
 (\alpha_0,W_0,b_0)=\arg\min_{\alpha,W,b}
 \mathbb E_{(x,y)\sim\mathcal D_t}\LCE(Wz_\alpha(x)+b,y),
 \label{eq:ce}
\end{equation}
which fixes the task representation before any margin refinement. For a true class $y$ and a competing class $j\ne y$, the signed Euclidean distance from $z_{\alpha_0}(x)$ to the boundary $\{z:(w_y-w_j)^\top z+b_y-b_j=0\}$ is
\begin{equation}
 m_{yj}(x)=\frac{(w_y-w_j)^\top z_{\alpha_0}(x)+(b_y-b_j)}{\|w_y-w_j\|_2}.
 \label{eq:margin}
\end{equation}
The decision region of class $y$ is an intersection of $K-1$ half-spaces, so for correctly classified $x$ the nearest-boundary margin $m(x)=\min_{j\ne y}m_{yj}(x)$ is exactly the distance from $z_{\alpha_0}(x)$ to the boundary of that region. Each $m_{yj}$ is $1$-Lipschitz in $z$, so a perturbation can shrink it by at most $\|r\|_2$, where $r=r_{\alpha_0}(x,\delta)$: $m_{yj}(x+\delta)\ge m_{yj}(x)-\|r\|_2$. If $\|r\|_2<m(x)$, every margin stays positive and the prediction does not change, which is Eq.~\eqref{eq:central}. Robustness therefore splits into two independently measurable quantities: movement, bounded by \Sone{} through Eq.~\eqref{eq:convexbound}, and margin, enlarged by \Stwo{}. The condition is sufficient, not necessary, so we use the two as directional diagnostics rather than as a predictor of RA.

\subsection{\Stwo: clean classifier-margin refinement}
Cross-entropy has no explicit control over the distance in Eq.~\eqref{eq:margin}: rescaling $W$ and $b$ raises every logit gap, and lowers the loss, without moving a boundary. \Stwo{} therefore optimizes that distance directly, and the normalization by $\|w_y-w_j\|_2$ makes it immune to the same rescaling. \Stwo{} starts from $(\alpha_0,W_0,b_0)$, freezes $\theta_r$ and $\alpha_0$, and updates only $W$ and $b$. Freezing the fusion is deliberate: margin optimization reshapes decision geometry without reselecting what clean adaptation chose, so Eq.~\eqref{eq:convexbound} still holds at $\alpha_0$. Because $m(x)$ is a non-smooth minimum, we optimize a soft nearest-boundary margin
\begin{equation}
 m_\tau(x)=-\tau\log\!\left[\sum_{j\ne y}
 \exp\!\left(-\frac{m_{yj}(x)}{\tau}\right)\right],
 \label{eq:softmargin}
\end{equation}
which satisfies $m(x)-\tau\log(K{-}1)\le m_\tau(x)\le m(x)$: \Stwo{} maximizes a lower bound on the true margin, exact as $\tau\to0$.
\begin{equation}
 (W^\star,b^\star)=\arg\min_{W,b}\mathbb E_{(x,y)\sim\mathcal D_t}\!\left[\LCE-\lambda_2m_\tau(x)\right],
 \label{eq:stage2}
\end{equation}
with $\lambda_2\ge0$. \Stwo{} never sees an adversarial example: $\lambda_2$, $\tau$, and the checkpoint come from clean training and validation data under a predeclared clean-accuracy constraint.

\section{Experimental Setup}
\label{sec:exp}

\begin{table*}[!t]
  \caption{
    Primary-backbone mechanism and validity suite.
    \textbf{(a)} \Sone{} variants with clean CE adaptation and without \Stwo{}.
    \textbf{(b)} CE, hard-margin, and soft-margin adaptation using the \Sone{} backbone.
    \textbf{(c)} Attack-wise and per-example-union RA for our full \Sone{}+\Stwo{} method.
    All values are averaged over three downstream adaptation seeds; overbars then denote four-task means of changes relative to Standard+CE.
  }
  \label{tab:abl}
  \vspace{1mm}

  \renewcommand{\arraystretch}{1.05}
  \setlength{\tabcolsep}{3.5pt}

  \noindent

  % ---------------------------------------------------------------
  % Panel (a)
  % ---------------------------------------------------------------
  {\scriptsize
  \begin{tabular}[t]{@{}lccc@{}}
    \multicolumn{4}{@{}c@{}}
      {\footnotesize\textbf{(a) \Sone{} variants (+CE)}} \\[1mm]
    \toprule

    % \textbf{Variant}
    %   & \shortstack{\textbf{$\overline{\Delta R_\alpha}$}\\
    %                 (\%) $\downarrow$}
    %   & \shortstack{\textbf{$\overline{\Delta\mathrm{CA}}$}\\
    %                 (pp) $\uparrow$}
    %   & \shortstack{\textbf{$\overline{\Delta\mathrm{RA}}$}\\
    %                 (pp) $\uparrow$} \\

    & $\overline{\Delta R_\alpha}$
    & $\overline{\Delta\mathrm{CA}}$
    & $\overline{\Delta\mathrm{RA}}$ \\
    % \noalign{\vskip -1pt}
    \textbf{Variant}
    & (\%) $\downarrow$
    & (pp) $\uparrow$
    & (pp) $\uparrow$ \\
    
    \midrule

    Terminal layer
      & -37.5
      & -3.0
      & 22.6 \\

    Mean all-layer
      & -72.9
      & -2.2
      & 39.3 \\

    STAGE 1, no $\Lpres$
      & -93.2
      & -20.8
      & 13.5 \\

    \rowcolor{oursbg}
    \textbf{\Sone{}}
      & -81.8
      & -2.1
      & 43.8 \\

    \bottomrule
  \end{tabular}}
  \hfill
  %
  % ---------------------------------------------------------------
  % Panel (b)
  % ---------------------------------------------------------------
  {\scriptsize
  \begin{tabular}[t]{@{}lccc@{}}
    \multicolumn{4}{@{}c@{}}
      {\footnotesize\textbf{(b) Margin refinement after \Sone{}}} \\[1mm]
    \toprule

    & $\overline{\Delta\mathrm{RA}}$
    & $\overline{\Delta M_{10}}$
    & $\overline{\Delta\mathrm{CA}}$ \\
    \noalign{\vskip 1pt}
    \textbf{Method}
    & (pp) $\uparrow$
    & ($\times10^{-3}$) $\uparrow$
    & (pp) $\uparrow$ \\
    
    \midrule

    CE
      & 43.8
      & 8.6
      & -2.1 \\

    hard margin
      & 45.5
      & 11.2
      & -2.9 \\

    \rowcolor{oursbg}
    \textbf{soft margin (\Stwo{})}
      & 48.0
      & 13.1
      & -3.2 \\

    \bottomrule
  \end{tabular}}
  \hfill
  %
  % ---------------------------------------------------------------
  % Panel (c)
  % ---------------------------------------------------------------
  {\scriptsize
  \begin{tabular}[t]{@{}lcccc@{}}
    \multicolumn{5}{@{}c@{}}
      {\footnotesize\textbf{(c) Full-method attack audit (RA)}} \\[1mm]
    \toprule

    \textbf{Attack}
      & \textbf{KS}
      & \textbf{IC}
      & \textbf{SID}
      & \textbf{ER} \\
    \midrule

    AutoPGD-CE
      & 69.5
      & 73.1
      & 40.5
      & 34.7 \\

    Normalized-margin (adaptive)
      & 68.4
      & 68.3
      & 42.2
      & 32.1 \\

    Fused-displacement (adaptive)
      & 76.7
      & 84.6
      & 51.8
      & 44.3 \\

    Psychoacoustic
      & 79.5
      & 86.3
      & 53.0
      & 45.8 \\

    \rowcolor{oursbg}
    \textbf{Per-example union}
      & 67.4
      & 66.7
      & 37.7
      & 29.6 \\

    \bottomrule
  \end{tabular}}

  \renewcommand{\arraystretch}{1.0}
\end{table*}

\textbf{Backbones and tasks.}
We evaluate Wav2Vec2 Large, HuBERT Large, and WavLM Large, which differ in pretraining objective and data~\cite{baevski2020wav2vec2,hsu2021hubert,chen2022wavlm}. We use Wav2Vec2 Large as the primary backbone for mechanism, attack, and sensitivity analyses, while all three backbones assess transfer breadth. We use four utterance-level tasks under the same frozen-backbone, learned-fusion, linear-head interface, spanning complementary speech factors: Speech Commands keyword spotting (KS; content)~\cite{warden2018speechcommands}, Fluent Speech Commands intent classification (IC; semantics)~\cite{lugosch2019fsc}, VoxCeleb1 speaker identification (SID; speaker)~\cite{nagrani2017voxceleb}, and four-class IEMOCAP emotion recognition (ER; paralinguistics)~\cite{busso2008iemocap,yang2021superb}. Comparisons are within backbone, not across differently pretrained models~\cite{yang2024largescale}.

\textbf{Threat model.}
Adversarial audio spans direct waveform, psychoacoustic, and physical attacks~\cite{carlini2018audio,yuan2018commandersong,schonherr2019psychoacoustic,qin2019imperceptible,yakura2019physical}; we focus on untargeted attacks with SNR-bounded digital waveform perturbations,
\begin{equation}
 \Dsnr(x)=\left\{\delta:\|\delta\|_2\le\|x\|_2 10^{-S/20},\;x+\delta\in\mathcal X\right\},
 \label{eq:threat}
\end{equation}
where $\mathcal X$ is the admissible sample range and norms are taken over valid unpadded samples. Headline results use $S=30$ dB, with $S\in\{24,27,30,33,36\}$ dB on the primary backbone for sensitivity analysis. The 30-dB operating point matches prior attacks on SSL speech recognizers~\cite{olivier2022watch}. Because global SNR is not a perceptual guarantee, we additionally include a psychoacoustically shaped speech-aware attack~\cite{schonherr2019psychoacoustic,qin2019imperceptible}.

\textbf{Robustness evaluation.}
Following standard adversarial evaluation practice~\cite{goodfellow2015adv,madry2018robust,athalye2018obfuscated}, we evaluate every model on the full test set using AutoPGD-CE~\cite{croce2020autoattack}, two defense-aware attacks that minimize the normalized classifier margin and maximize fused-representation displacement, respectively, and the psychoacoustically shaped attack. All four attacks run for 50 steps with 2 restarts, satisfy the 30 dB SNR constraint, and differentiate through preprocessing, pooling, learned fusion, and classifier. RA is the fraction of examples that remain correct under \emph{every attack in the pool}, i.e., the per-example union rather than the minimum of separately averaged accuracies. To quantify the movement--margin mechanism, Table~\ref{tab:main} reports each method's changes relative to its matched Standard+CE baseline. Let $\bar R_\alpha=\mathbb E_x\|r_\alpha(x,\delta_x)\|_2$ denote the mean fused-representation movement, where $\delta_x$ is the attack in the pool that produces the largest displacement, computed against the model being measured. We define
\begin{equation}
\Delta R_\alpha=\frac{\bar R_\alpha-\bar R_\alpha^{\mathrm{Std{+}CE}}}
                     {\bar R_\alpha^{\mathrm{Std{+}CE}}},
\qquad
\Delta M_{10}=m_{10}-m_{10}^{\mathrm{Std{+}CE}},
\label{eq:diagnostic_changes}
\end{equation}
where $m_{10}$ is the 10th percentile of the clean normalized margin. Negative $\Delta R_\alpha$ indicates reduced adversarial representation movement, whereas positive $\Delta M_{10}$ indicates a larger lower-tail margin. On the primary backbone we verify the achieved SNR and confirm that RA no longer decreases beyond 50 steps and 2 restarts.

\textbf{Baselines and controls.}
Table~\ref{tab:main} compares Standard+CE, FARE-style robustification~\cite{schlarmann2024robustclip}, a matched \emph{Terminal-layer} control, our two-stage method, and downstream adversarial references. FARE-style minimizes one term, the displacement of the attacked final-layer output from its clean value under $\theta_b$, so robustness and preservation are not separated; it uses the same unlabeled data and 30-dB threat as \Sone{}. Terminal-layer instead keeps \Sone{}'s $\Lpres$ unchanged and replaces only $\operatorname{LSE}_\beta(\mathbf d)$ by $d_L$. Adversarial head training keeps the \Sone{} backbone and learned fusion fixed while training the classifier on downstream adversarial examples; full AT additionally updates the backbone and is reported only as a primary-backbone oracle.

\textbf{Training, selection, and reporting.}
\Sone{} uses 100\,h of Common Voice~\cite{ardila2020commonvoice} as unlabeled robustification data and is trained for three epochs with $\lambda_1=0.2$, 10-step PGD at 30\,dB, and $\beta=30$, which causes $\operatorname{LSE}_\beta$ to emphasize higher-drift layers, especially at the start of robustification; $\sigma_\ell$ is calibrated on a disjoint held-out split. \Stwo{} uses $\lambda_2=2$, $\tau=0.1$, and trains for 50 epochs on cached fused representations. Both stages use Adam with 10\% linear warm-up and linear decay to zero, with peak learning rates $5\times10^{-5}$ for \Sone{} and $10^{-3}$ for \Stwo{}. \Stwo{} hyperparameters are selected on held-out clean data; no adversarial validation or test examples are used. Tables~\ref{tab:main}--\ref{tab:abl} report values across three downstream seeds for a fixed \Sone{} checkpoint; $\Delta$ quantities are paired by seed and overbars denote four-task means.

\textbf{Mechanism and validity audits.}
On the primary backbone, Table~\ref{tab:abl}(a) ablates \Sone{}'s across-depth aggregation, worst-layer weighting, and clean anchor. Panel~(b) compares CE, hard-margin, and soft-margin objectives with the backbone and fusion fixed. Panel~(c) reports attack-wise and per-example-union RA. Fig.~\ref{fig:layerdrift} shows layer-wise drift and the 24--36\,dB SNR sweep.

\section{Results and Analysis}
\label{sec:results}

\textbf{Robustness transfer.}
\Sone{}+CE improves RA over Standard+CE on all 12 backbone--task pairs, by 46.4\,pp
on average at a 2.1-pp CA cost, and exceeds the Terminal-layer control by 21.7\,pp
RA at 0.9\,pp higher CA (Table~\ref{tab:main}). Stabilizing the hidden hierarchy
before the task is known therefore transfers robustness that terminal-layer
stabilization does not.

\textbf{Clean margin refinement.}
\Stwo{} adds RA on every pair, 4.0\,pp on average for 1.1\,pp CA, using no
downstream adversarial examples; the soft surrogate beats the hard minimum by
2.5\,pp RA with a larger $m_{10}$ (Table~\ref{tab:abl}(b)). Using downstream adversarial examples, head training gains 2.7\,pp more RA for 2.1\,pp less CA; full AT reaches 63.0\% mean RA against our 50.4\% on the primary backbone, but updates the backbone per task.

\begin{figure}[t]
\centering
\includegraphics[width=\columnwidth]{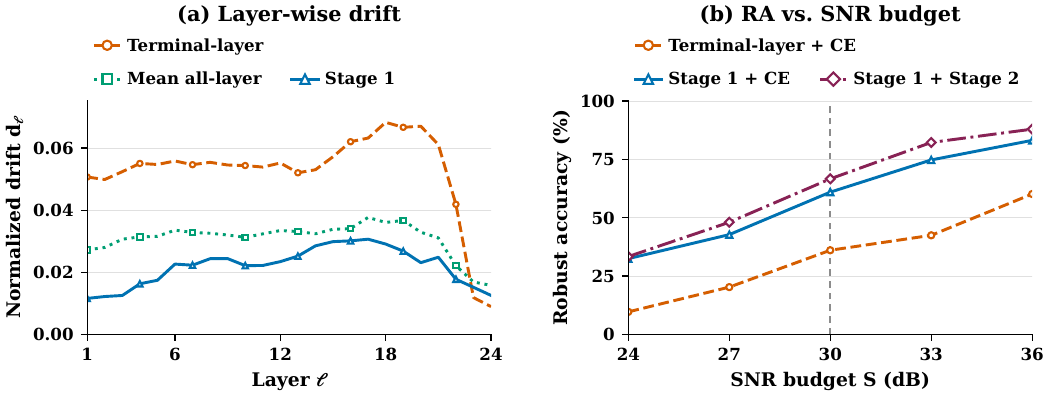}\vspace{-1mm}
\caption{Primary-backbone diagnostics on the IC task. \textbf{(a)} Layer-wise representation drift under terminal-layer, mean all-layer, and \Sone{} stabilization. \textbf{(b)} Robust accuracy across 24--36\,dB SNR budgets.}
\label{fig:layerdrift}
\vspace{-1.5mm}
\end{figure}

\textbf{Hierarchy and preservation.}
\Sone{} exceeds mean all-layer stabilization by 4.5\,pp RA (Table~\ref{tab:abl}(a)),
and Fig.~\ref{fig:layerdrift}(a) shows why: the terminal-layer objective attains the
lowest drift at layer $L$ but the highest drift at most earlier depths, while worst-layer
weighting pushes \Sone{}'s peak drift below the layer-mean variant. Removing $\Lpres$ gives the largest movement reduction (93.2\%)
yet costs 18.7\,pp CA and 30.3\,pp RA against \Sone{}:
reduced movement alone is not robustness.

\textbf{Adaptive evaluation.}
The normalized-margin attack is strongest on KS, IC, and ER and AutoPGD-CE on SID,
and the union removes a further 1.0--2.8\,pp, so the attacks fail on different
examples (Table~\ref{tab:abl}(c)). The method ranking is unchanged across
24--36\,dB (Fig.~\ref{fig:layerdrift}(b)).

\textbf{Training cost.}
On an RTX~4090, \Sone{} costs 18.9 GPU-h once per backbone, while \Stwo{} costs 0.37 GPU-h per task, compared with 10.7 GPU-h per task for full AT.

\section{Conclusion}
\label{sec:conclusion}
Stabilizing the hidden hierarchy before tasks are known transfers
robustness beyond terminal-layer stabilization; clean margin refinement
further improves robustness without downstream adversarial examples.
Across 12 backbone--task pairs, the full method improves mean RA
by 50.4\,pp over Standard+CE at a 3.2\,pp CA cost under the
evaluated digital waveform attacks.

\clearpage

\noindent\textbf{Acknowledgment.} Mohammad Sabokrou was supported by JSPS KAKENHI Grant-in-Aid for Scientific Research (B), Grant Number JP26K02988.

\vspace{1mm}

\noindent\textbf{Compliance with Ethical Standards.} This research was conducted retrospectively on existing human speech corpora (Common Voice, Speech Commands, Fluent Speech Commands, VoxCeleb1, IEMOCAP), each obtained and used under its respective license.

\vspace{1mm}

\let\OLDthebibliography\thebibliography
\renewcommand\thebibliography[1]{%
  \OLDthebibliography{#1}%
  \setlength{\parskip}{0pt}%
  \setlength{\itemsep}{0pt plus 0.2ex}%
}
\bibliographystyle{IEEEbib}
\bibliography{refs}

@inproceedings{baevski2020wav2vec2,
  author    = {Alexei Baevski and Henry Zhou and Abdelrahman Mohamed and Michael Auli},
  title     = {{wav2vec 2.0}: A Framework for Self-Supervised Learning of Speech Representations},
  booktitle = {NeurIPS},
  volume    = {33},
  year      = {2020}
}

@article{hsu2021hubert,
  author  = {Wei-Ning Hsu and Benjamin Bolte and Yao-Hung Hubert Tsai and others},
  title   = {{HuBERT}: Self-Supervised Speech Representation Learning by Masked Prediction of Hidden Units},
  journal = {IEEE/ACM TASLP},
  volume  = {29},
  pages   = {3451--3460},
  year    = {2021}
}

@article{chen2022wavlm,
  author  = {Sanyuan Chen and Chengyi Wang and Zhengyang Chen and Yu Wu and Shujie Liu and others},
  title   = {{WavLM}: Large-Scale Self-Supervised Pre-Training for Full Stack Speech Processing},
  journal = {IEEE JSTSP},
  volume  = {16},
  number  = {6},
  pages   = {1505--1518},
  year    = {2022},
  doi     = {10.1109/JSTSP.2022.3188113}
}

@inproceedings{pasad2023layerwise,
  author    = {Ankita Pasad and Bowen Shi and Karen Livescu},
  title     = {Comparative Layer-Wise Analysis of Self-Supervised Speech Models},
  booktitle = {Proc. ICASSP},
  pages     = {1--5},
  year      = {2023},
  doi       = {10.1109/ICASSP49357.2023.10096149}
}

@inproceedings{yang2021superb,
  author    = {Shu-wen Yang and Po-Han Chi and Yung-Sung Chuang and Cheng-I Jeff Lai and others},
  title     = {{SUPERB}: Speech Processing Universal {PER}formance Benchmark},
  booktitle = {Proc. Interspeech},
  pages     = {1194--1198},
  year      = {2021},
  doi       = {10.21437/Interspeech.2021-1775}
}

@article{yang2024largescale,
  author  = {Shu-Wen Yang and Heng-Jui Chang and Zili Huang and Andy T. Liu and others},
  title   = {A Large-Scale Evaluation of Speech Foundation Models},
  journal = {IEEE/ACM TASLP},
  volume  = {32},
  pages   = {2884--2899},
  year    = {2024},
  doi     = {10.1109/TASLP.2024.3389631}
}

@article{warden2018speechcommands,
  author  = {Pete Warden},
  title   = {Speech Commands: A Dataset for Limited-Vocabulary Speech Recognition},
  journal = {arXiv preprint arXiv:1804.03209},
  year    = {2018}
}

@inproceedings{lugosch2019fsc,
  author    = {Loren Lugosch and Mirco Ravanelli and Patrick Ignoto and others},
  title     = {Speech Model Pre-Training for End-to-End Spoken Language Understanding},
  booktitle = {Proc. Interspeech},
  pages     = {814--818},
  year      = {2019},
  doi       = {10.21437/Interspeech.2019-2396}
}

@inproceedings{nagrani2017voxceleb,
  author    = {Arsha Nagrani and Joon Son Chung and Andrew Zisserman},
  title     = {{VoxCeleb}: A Large-Scale Speaker Identification Dataset},
  booktitle = {Proc. Interspeech},
  pages     = {2616--2620},
  year      = {2017},
  doi       = {10.21437/Interspeech.2017-950}
}

@article{busso2008iemocap,
  author  = {Carlos Busso and Murtaza Bulut and Chi-Chun Lee and Abe Kazemzadeh and Emily Mower and others},
  title   = {{IEMOCAP}: Interactive Emotional Dyadic Motion Capture Database},
  journal = {Lang. Resour. Eval.},
  volume  = {42},
  number  = {4},
  pages   = {335--359},
  year    = {2008},
  doi     = {10.1007/s10579-008-9076-6}
}

@inproceedings{shafahi2020robusttransfer,
  author    = {Ali Shafahi and Parsa Saadatpanah and Chen Zhu and Amin Ghiasi and Christoph Studer and others},
  title     = {Adversarially Robust Transfer Learning},
  booktitle = {ICLR},
  year      = {2020}
}

@inproceedings{chen2020advssl,
  author    = {Tianlong Chen and Sijia Liu and Shiyu Chang and others},
  title     = {Adversarial Robustness: From Self-Supervised Pre-Training to Fine-Tuning},
  booktitle = {Proc. CVPR},
  pages     = {699--708},
  year      = {2020}
}

@inproceedings{nern2023transfer,
  author    = {Laura F. Nern and Harsh Raj and Maurice Andr\'e Georgi and Yash Sharma},
  title     = {On Transfer of Adversarial Robustness from Pretraining to Downstream Tasks},
  booktitle = {NeurIPS},
  volume    = {36},
  year      = {2023},
  doi       = {10.52202/075280-2584}
}

@inproceedings{hua2024roli,
  author    = {Andong Hua and Jindong Gu and Zhiyu Xue and others},
  title     = {Initialization Matters for Adversarial Transfer Learning},
  booktitle = {Proc. CVPR},
  pages     = {24831--24840},
  year      = {2024}
}

@inproceedings{radford2021clip,
  author    = {Alec Radford and Jong Wook Kim and Chris Hallacy and Aditya Ramesh and others},
  title     = {Learning Transferable Visual Models From Natural Language Supervision},
  booktitle = {Proc. ICML},
  volume    = {139},
  pages     = {8748--8763},
  year      = {2021}
}

@inproceedings{mao2023tecoa,
  author    = {Chengzhi Mao and Scott Geng and Junfeng Yang and others},
  title     = {Understanding Zero-Shot Adversarial Robustness for Large-Scale Models},
  booktitle = {ICLR},
  year      = {2023}
}

@inproceedings{schlarmann2024robustclip,
  author    = {Christian Schlarmann and Naman Deep Singh and Francesco Croce and Matthias Hein},
  title     = {Robust {CLIP}: Unsupervised Adversarial Fine-Tuning of Vision Embeddings for Robust Large Vision-Language Models},
  booktitle = {Proc. ICML},
  volume    = {235},
  pages     = {43685--43704},
  year      = {2024}
}

@inproceedings{goodfellow2015adv,
  author    = {Ian J. Goodfellow and Jonathon Shlens and Christian Szegedy},
  title     = {Explaining and Harnessing Adversarial Examples},
  booktitle = {ICLR},
  year      = {2015}
}

@inproceedings{madry2018robust,
  author    = {Aleksander Madry and Aleksandar Makelov and Ludwig Schmidt and others},
  title     = {Towards Deep Learning Models Resistant to Adversarial Attacks},
  booktitle = {ICLR},
  year      = {2018}
}

@inproceedings{athalye2018obfuscated,
  author    = {Anish Athalye and Nicholas Carlini and David Wagner},
  title     = {Obfuscated Gradients Give a False Sense of Security: Circumventing Defenses to Adversarial Examples},
  booktitle = {Proc. ICML},
  volume    = {80},
  pages     = {274--283},
  year      = {2018}
}

@inproceedings{croce2020autoattack,
  author    = {Francesco Croce and Matthias Hein},
  title     = {Reliable Evaluation of Adversarial Robustness with an Ensemble of Diverse Parameter-Free Attacks},
  booktitle = {Proc. ICML},
  volume    = {119},
  pages     = {2206--2216},
  year      = {2020}
}

@inproceedings{carlini2018audio,
  author    = {Nicholas Carlini and David Wagner},
  title     = {Audio Adversarial Examples: Targeted Attacks on Speech-to-Text},
  booktitle = {IEEE S\&P Workshops},
  pages     = {1--7},
  year      = {2018},
  doi       = {10.1109/SPW.2018.00009}
}

@inproceedings{yuan2018commandersong,
  author    = {Xuejing Yuan and Yuxuan Chen and Yue Zhao and Yunhui Long and Xiaokang Liu and others},
  title     = {{CommanderSong}: A Systematic Approach for Practical Adversarial Voice Recognition},
  booktitle = {USENIX Security},
  pages     = {49--64},
  year      = {2018}
}

@inproceedings{schonherr2019psychoacoustic,
  author    = {Lea Sch\"onherr and Katharina Kohls and Steffen Zeiler and others},
  title     = {Adversarial Attacks Against Automatic Speech Recognition Systems via Psychoacoustic Hiding},
  booktitle = {NDSS},
  year      = {2019},
  doi       = {10.14722/ndss.2019.23288}
}

@inproceedings{qin2019imperceptible,
  author    = {Yao Qin and Nicholas Carlini and Garrison W. Cottrell and others},
  title     = {Imperceptible, Robust, and Targeted Adversarial Examples for Automatic Speech Recognition},
  booktitle = {Proc. ICML},
  volume    = {97},
  pages     = {5231--5240},
  year      = {2019}
}

@inproceedings{yakura2019physical,
  author    = {Hiromu Yakura and Jun Sakuma},
  title     = {Robust Audio Adversarial Example for a Physical Attack},
  booktitle = {Proc. IJCAI},
  pages     = {5334--5341},
  year      = {2019},
  doi       = {10.24963/ijcai.2019/741}
}

@inproceedings{wu2022sslattack,
  author    = {Haibin Wu and Bo Zheng and Xu Li and others},
  title     = {Characterizing the Adversarial Vulnerability of Speech Self-Supervised Learning},
  booktitle = {Proc. ICASSP},
  pages     = {3164--3168},
  year      = {2022},
  doi       = {10.1109/ICASSP43922.2022.9747242}
}

@article{olivier2022watch,
  author  = {Rapha\"el Olivier and Hadi Abdullah and Bhiksha Raj},
  title   = {Watch What You Pretrain For: Targeted, Transferable Adversarial Examples on Self-Supervised Speech Recognition Models},
  journal = {arXiv preprint arXiv:2209.13523},
  year    = {2022}
}

@inproceedings{wang2026vac,
  author    = {Yanyun Wang and Baoyuan Wu and Li Liu},
  title     = {Adversarial Fine-Tuning on Speech Foundation Model with Vulnerable Attention Consistency Regularization for Robust Speech Recognition},
  booktitle = {Proc. ICASSP},
  year      = {2026},
  doi       = {10.1109/ICASSP55912.2026.11462855}
}

@inproceedings{ardila2020commonvoice,
  author    = {Rosana Ardila and Megan Branson and Kelly Davis and others},
  title     = {{Common Voice}: A Massively-Multilingual Speech Corpus},
  booktitle = {Proc. LREC},
  pages     = {4218--4222},
  year      = {2020}
}

\end{document}